\documentclass[aps,pra, reprint,superscriptaddress,showkeys]{revtex4-2}
\usepackage{graphicx}
\usepackage{units} 
\usepackage{lineno}
\usepackage{subfigure}
\usepackage[table]{xcolor}
\usepackage{amsmath}

\setkeys{Gin}{width=1.0\linewidth} 

\begin{document}
\preprint{Version 2}

\author{Vladimir M. Subbotin}
\email[Email:]{vsubbotin@arrowheadpharma.com}
\affiliation{Arrowhead Pharmaceuticals Inc., Madison, WI, USA}
\affiliation{University of Wisconsin, Madison, WI, USA}

\author{Benjamin A. Turner}
\affiliation{Arrowhead Pharmaceuticals Inc., Madison, WI, USA}

\author{Brian A. Davies}
\affiliation{Arrowhead Pharmaceuticals Inc., Madison, WI, USA}

\author{Alric G. Lopez}
\affiliation{Arrowhead Pharmaceuticals Inc., Madison, WI, USA}

\author{Gennady Fiksel}
\affiliation{University of Wisconsin, Madison, WI, USA}

\title{Experimental and numerical modeling of liposome congregation in meteorite craters of Early Earth\thanks{Parts of this material were presented at the European Astrobiology 
Conference ``Impacts and their Role in the Origin and Evolution of Life'', N\"{o}rdlingen, Germany, September 1-4, 2025.}} 





\begin{abstract}
This paper provides experimental and numerical evidence supporting the occurrence of liposome congregation at the floors of meteor craters on Early Earth. 
This work builds on our earlier research, which demonstrated that liposomes submerged in a shallow Archean pond are protected from harmful UV radiation.
This protection allows them to survive long enough for autocatalytic replication of amphiphiles and for mutation and selection of assemblies that maximize membrane stability. 
For liposomes to fuse, grow, exchange contents and membrane components, and divide, they need to establish a population, which means forming a dense conglomerate that enables close physical contact.
The study demonstrates that such a congregation is feasible in bowl-shaped meteor craters on Early Earth, especially under periodic seismic disturbances.
\end{abstract}


\maketitle 

\section{Introduction}
In our earlier publication~\cite{Subbotin2023AstrobiologyLipidWorld}, we proposed a hypothesis on the Darwinian evolution of liposomes based solely on 
natural and ever-present phenomena: the diurnal cycle of solar UV radiation, gravity, and the formation and release of amphiphiles in aqueous media.
The hypothesis is based on the premise that amphiphilic molecules introduced into Archean water inevitably accumulate at the water-air interface and form Langmuir layers, bilayers, and liposomes.

In our scenario, the newly formed liposomes would inevitably be destroyed by solar UV unless they acquire negative buoyancy by capturing heavy solutes, such as ribose, and descend from the water-air boundary. 
We have also shown that some primordial water constituents, such as ferric salts, can provide UV attenuation sufficient to shield liposomes 
at depths of only a few millimeters~\cite{Subbotin2023AstrobiologyUVProtection,Turner2025FrontiersUVLiposomes}. 
The submerged liposomes, shielded from damaging UV radiation, acquire the longevity necessary 
to enhance autocatalytic replication of amphiphiles, their mutation, and the selection of those amphiphilic assemblies that provide the greatest membrane stability. 
These two types of mutable and heritable compositional information-heavy content and amphiphilic assembly design-constitute adaptive traits influenced by natural selection. 
Selection and propagation of the best-fitted traits constitute Darwinian evolution.

However, liposome survival alone, while necessary, is not sufficient to be a subject of Darwinian evolution.
For that, the surviving liposomes must form a population~\cite{Adamski2020NatRevChemReplication,Mariano2024LifeGameTheory,Okasha2022FrontiersTransitions, Takeuchi2009PLOSCompBioMultilevel}. 
It is worth noting that although Charles Darwin did not use the term  ``population'',
his book ``The Origin of Species ...'' \cite{Darwin1859origin} undoubtedly contains the concept of a population.
For example, in Chapter II, p.45 and Chapter IV, p.84, Darwin writes: 
``These individual differences are highly important for us, as they afford materials for natural selection to accumulate, 
it may be said that natural selection is daily and hourly scrutinising, throughout the world, every variation, even the slightest; rejecting that which is bad, preserving and adding up all that is good.''
The words ``adding up'' and ``individual differences'' make sense only at the population level, not at the level of single individuals. 
In addition, contemporary analyses also emphasize that a population, or the formation of groups of similar entities, such as protocells, arising through temporal and spatial congregation, is fundamental for the Origin of Life~\cite{Szathmary1987JTheorBiolGroupSelection}. 
Therefore, for the liposomes to be subject to Darwinian evolution, they must congregate and form a population.

Multiple in-vitro studies showed that liposomes, when subjected to physical contact with each other, undergo fusion and growth~\cite{Connor1984PNASpHLiposomes,Deshpande2019SmallGrowth,Nir1983ProgSurfSciVesicles,Noguchi2001JCPFusion}, exchange of liposomal content and membrane components~\cite{Chan2008BiointerphasesDNAFusion,Hanczyc2003ScienceCompartments,Hardy2015PNASPhospholipids,Yang2009SmallNanoparticles}, and fusion and division~\cite{Deshpande2018ACSNanoDivision,Dobereiner1993BiophysJVesicles,Penic2020FrontiersPhysicsBudding}.
The issue we are trying to address in this study is: how might such a liposomal population be formed if liposomes are just scattered on the floor of primordial ponds?

Various studies~\cite{Deamer1989OLEBMMurchison,Dworkin2001PNASAmphiphiles,Martins2024ElementsDelivery,Zhao2025GCAPhospholipidImpact} 
demonstrated that sites of meteorite impacts could be seeded with amphiphiles, phospholipid precursors, and ribose~\cite{Abe2024ACSEarthSugars,Furukawa2019PNASRibose,Ono2024AstrobiologyRibose,Paschek2022LifeRibose}, the molecules that are essential for our hypothesized origin of life~\cite{Subbotin2023AstrobiologyLipidWorld}. 
More so, a recent experimental study demonstrated that the meteorite impacts themselves produce phospholipid precursors directly from Earth's soil, in both wet and dry conditions~\cite{Zhao2025GCAPhospholipidImpact}.
Therefore, it is plausible to consider that meteorite craters on Hadean Earth could have been potential sites for the self-sustained Darwinian evolution of liposomes~\cite{Cockell2006PhilTransImpact,Osinski2020AstrobiologyImpact}.

But what about forming a population? Meteorite craters, as shown by surface images of the Moon and Mars, exhibit a wide range of shapes and sizes~\cite{Osinski2020AstrobiologyImpact}.
Generally, their topography consists of concave, bowl-shaped depressions, dips, and crevices at scales ranging from tens of kilometers to meters and less.
Each of these features, filled with water enriched with various minerals, particularly ferric salts, can serve as a site for liposome formation, survival, and growth.
Although the concave shape of the sites theoretically encourages heavy liposomes to settle at the bottom by minimizing gravitational potential energy, this process is often impeded by friction and physical obstacles.
However, external disturbances, such as local earthquakes and shock waves produced by meteorite bombardment, may shift the soil, thereby releasing liposomes and facilitating their downward migration. 
To explore this problem, we developed experimental and numerical models of liposomal dynamics within meteorite craters subjected to sudden, repetitive displacements.

The paper is organized as follows. 
Section~\ref{sec:Experiment} describes the experimental setup and results.
Section~\ref{sec:Numerics} describes numerical simulations of the experiments.
Section~\ref{sec:UV_Crater} illustrates how liposomes are protected against UV damage in the crater environment.
Section~\ref{sec:Summary} summarizes the results.

\section{Experiment}
\label{sec:Experiment}

\subsection{Earthquake literature data}
\label{subsec:Earthquake literature data}
The experiment described in this section is designed and built to closely simulate
the ground motion characteristics observed during earthquakes, such as the soil displacement magnitude and velocity.
Detailed earthquake data, including amplitude and frequency spectra of various tremor types, can be found in~\cite{Mohraz2001}.
Displacement measurements, assembled from data across multiple earthquake sites, vary from centimeters to tens of centimeters, while velocity values vary from a fraction of $\unit [1]{m/s}$ to several~ m/s. The experiment and numerical simulations are designed to match these characteristics.

\subsection{Experimental setup }
\label{sec:Setup}

\begin{figure}[htbp]
\begin{center}
\includegraphics{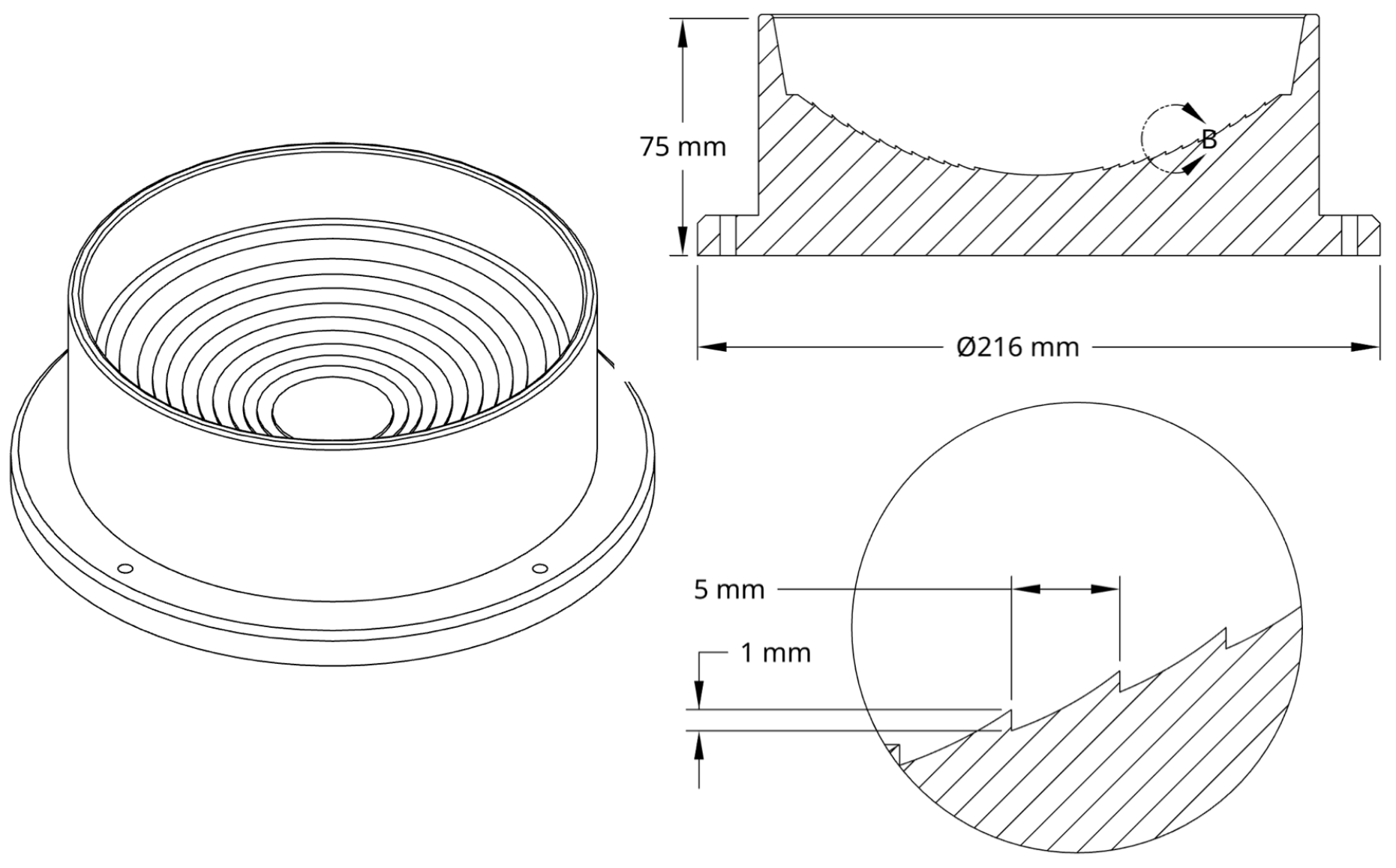} 
\caption{A drawing of a crater model showing an overview and detail views.}
\label{fig:Crater}
\end{center}
\end{figure}

A ``crater'' model drawing is shown in Fig.~\ref{fig:Crater}.
The crater has a spherical cap shape with a sphere radius of $\unit [200]{mm}$, a cap radius of $\unit [80]{mm}$, and a height of $\unit [35]{mm}$.
It is produced using water-tight 3D printing technology.
The base of the crater features ridges spaced every $\unit [5]{mm}$, each measuring $\unit [1]{mm}$ in height.
The crater is filled with water into which polyethylene red liposomes-mimicking microspheres~\cite{Cospheric}, 
with a diameter of $d = \unit [350]{\mu m}$ and density of $\rho = \unit [1.2]{g/cm^{3}}$ are submerged. 
The purpose of the ridges is to provide physical obstacles preventing the microspheres from rolling down the hill under gravity.
Clearly, neither the size and shape of the ``crater" nor the size, material, and mass of the ``liposomes'' approach their real-life counterparts.
Rather, the purpose of this experiment is a demonstration of a general principle of liquid-submerged micro-particulate dynamics, driven
by gravitation and assisted by periodic disturbances.
In addition, comparing the experimental and numerical results will validate the numerical model and provide a basis for further development and improvement.

\begin{figure}[htbp]
\begin{center}
\includegraphics[width=0.75\linewidth]{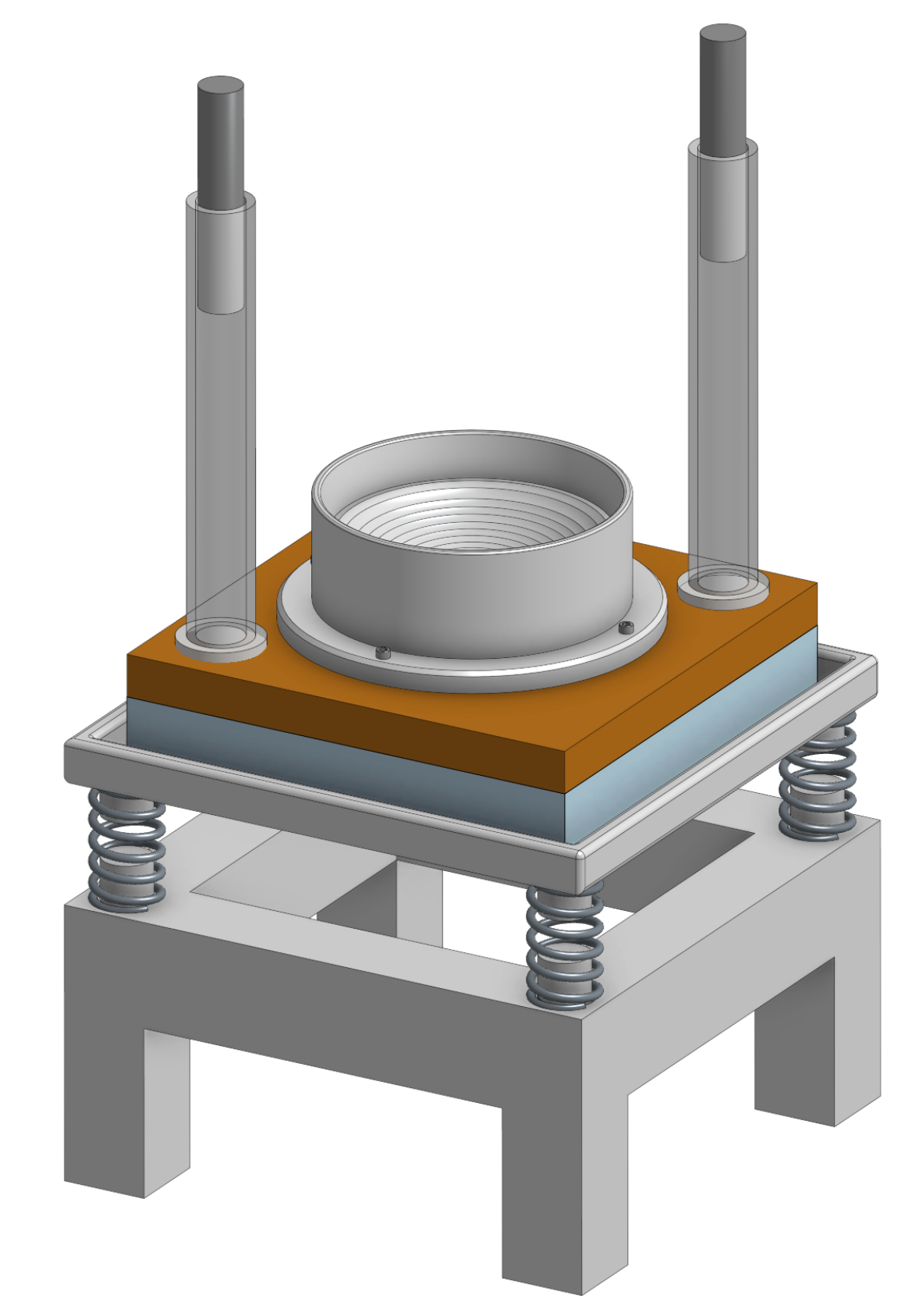} 
\caption{Experimental setup.
A stainless steel tray is connected to a stand with four springs. 
The crater model is mounted on a $\unit [2]{cm}$-thick wooden plate, which is placed on top of a $\unit [3]{cm}$-thick polyurethane foam layer resting within the tray.
Two weights are dropped from $h = \unit [25] {cm}$.}
\label{fig:table}
\end{center}
\end{figure}

An experimental setup for investigating liposome behavior is shown in Fig.~\ref{fig:table} and consists of a stainless steel tray connected to a stand with four springs.
The crater model is mounted on a $\unit [2]{cm}$-thick wooden plate, which is placed on top of a $\unit [3]{cm}$-thick polyurethane foam layer resting within the tray.
Two weights are dropped from  $ h=\unit [25] {cm}$, thereby generating the crater's displacement through momentum transfer, thus emulating an initial quake jolt.

Since the foam offers minimal resistance at the initial moment, the crater velocity at impact can be estimated using momentum conservation

\begin{equation}
\label{eg:MomentumBalance}
V_c = V_w \frac {M_w} {M_w+M_c+M_b},
\end{equation}
where $M_w$, $M_c$, and $M_b$ are the masses of the drop weights, the crater, and the board, respectively, and $V_w$ is the drop weight velocity at the impact.

As measured, $M_c + M_b= \unit [1.9] {kg}$  and a calculated value of  $V_w = \sqrt{2gh} = \unit [2.2] {m/s}$.
Two kinds of drop weights were used - steel bars with a total mass of $M_w = \unit [2.4]{kg}$, and aluminum bars with a total mass of $M_w = \unit [0.83]{kg}$. 
Applying Eq.~\ref{eg:MomentumBalance} results in the velocity of the crater at the impact being $V_c = \unit [1.2] {m/s}$ and $V_c = \unit [0.67] {m/s}$, respectively.
The impact velocity figures align well with the earthquake data presented in Section~\ref{subsec:Earthquake literature data}.

\subsection{Experimental results}
\begin{figure}[htbp]
\begin{center}
\includegraphics{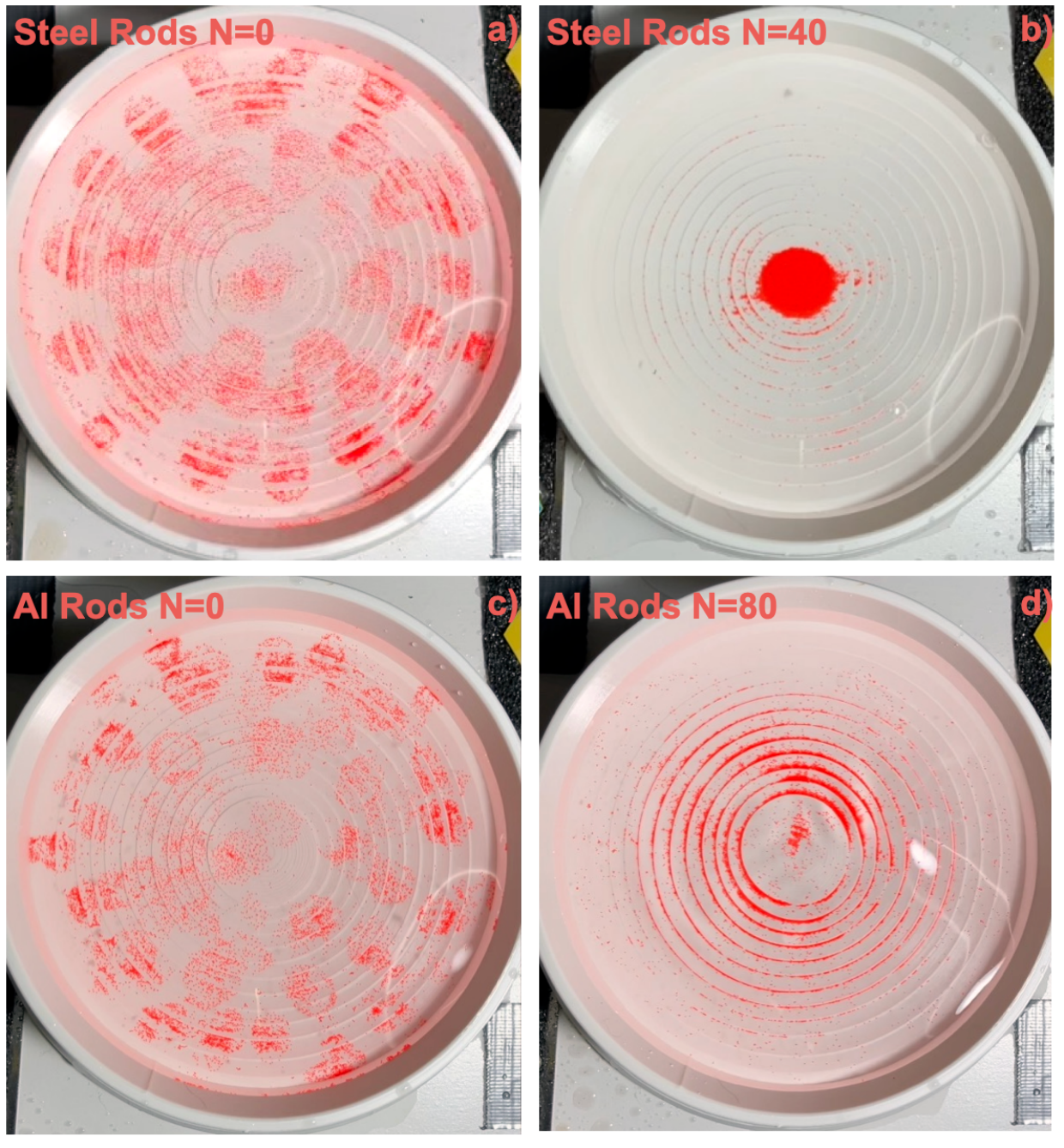} 
\caption{Summary of particle dynamics.
The top and bottom rows compare results for steel and aluminum rods, respectively, while the left and right columns show the initial and final stages.
(b) Dropping steel rods leads to nearly complete particle collection at the center after N = 40 drops. 
(d) Dropping aluminum rods results in a profile that, although somewhat peaked, remains noticeably wide even after 80 drops.}
\label{fig:Steel_vs_Al}
\end{center}
\end{figure}

Before each test, the red microspheres were spread evenly across the entire crater. 
Afterwards, a series of repeated weight drops was performed, and the movement of the particles was recorded on video.
The results are summarized in Fig.~\ref{fig:Steel_vs_Al}.
The top and bottom rows compare results for steel and aluminum rods, respectively, while the left and right columns show the initial and final stages.
Dropping steel rods leads to nearly complete particle collection at the center after N = 40 drops. 
Dropping aluminum rods results in a profile that, although somewhat peaked, remains noticeably wide even after 80 drops.

\section{Numerical simulations }
\label{sec:Numerics}

\begin{figure}[htbp]
\begin{center}
\includegraphics{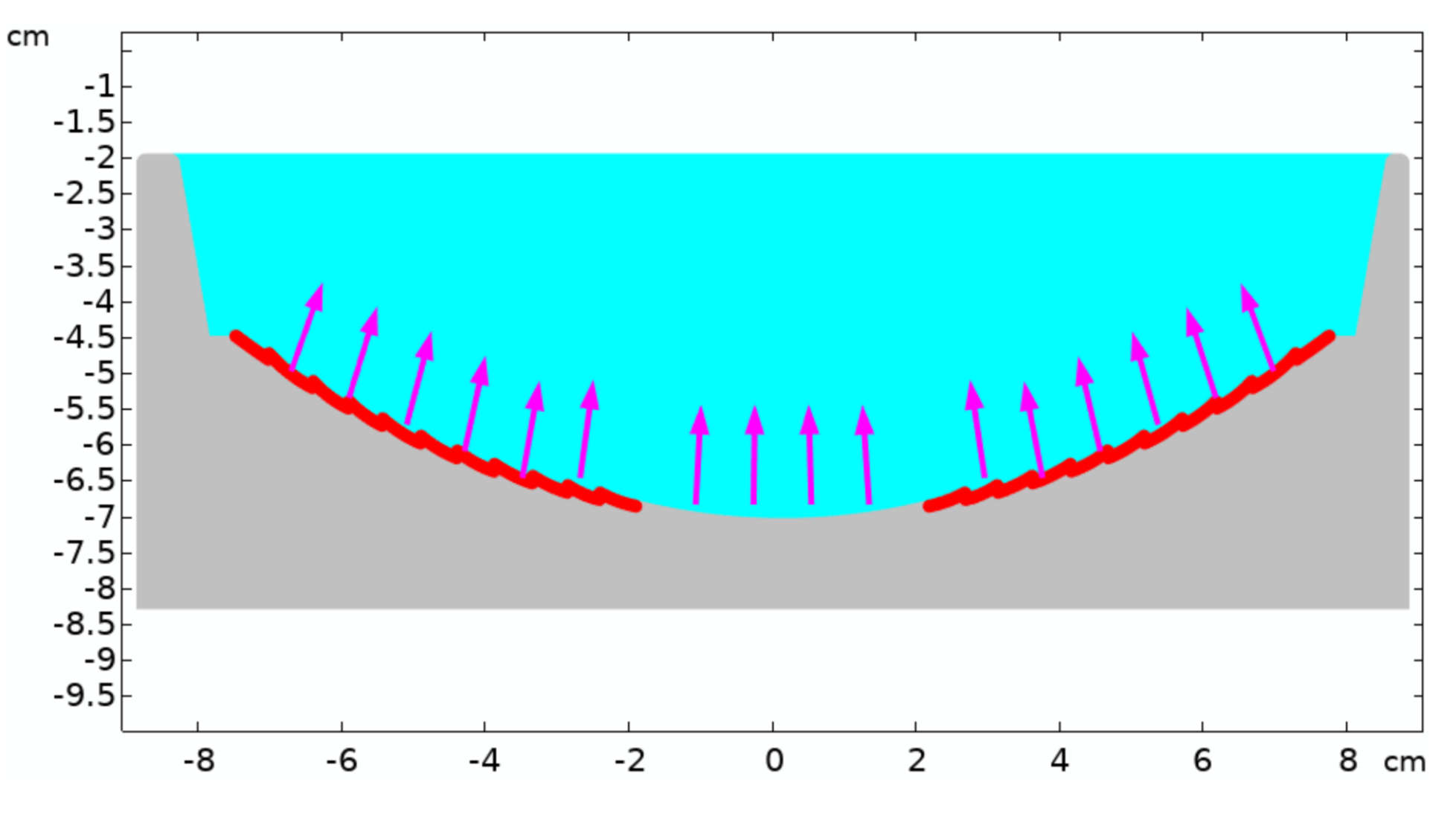} 
\caption{A thousand particles, each with a diameter of $d = \unit [350]{\mu m}$ and density of $\rho = \unit [1.2]{g/cm^3}$ are submerged in water (colored in cyan) 
and are initially uniformly distributed across the curved bottom, as shown in the red circles.
It is assumed that after the weights impact the crater, the crater descends and then rebounds, striking the particles and initiating their motion. 
Because of the concave shape of the bottom, the momentum transferred to the particles has both vertical and horizontal components, depending on their initial positions, as indicated by the arrows.
}
\label{fig:Simulations_T0}
\end{center}
\end{figure}

The behavior of the microspheres (called particles in the code) in the experiment is modeled using COMSOL$\textsuperscript{\textregistered}$ Multiphysics software~\cite{COMSOLMultiphysicsReference}.
The simulation geometry shown in Fig.~\ref{fig:Simulations_T0} duplicates the experimental model exactly, except that the simulations are conducted in 2D.

A thousand ''numerical'' particles, each with a diameter of $d = \unit [350]{\mu m}$ and density of $\rho = \unit [1.2]{g/cm^3}$ are submerged in water and 
are initially uniformly distributed across the curved bottom, as shown in the red circles.
It is assumed that after the weights impact the crater,  the crater descends and then rebounds, striking the particles and initiating their motion.
Because of the concave shape of the bottom, the momentum transmitted to the particles will have both vertical and horizontal components, depending on their initial positions, as indicated by the arrows.
Specifically, the horizontal velocity component of each particle is  $V_0\sin(2\alpha)$ and the vertical component is $V_0(1+\cos(2\alpha))$, where $V_0$ is the crater rebound velocity
and $\sin \alpha = x/R$, where $x$ is the particle horizontal position at the impact and $R$ is the radius of the concave cavity. 
The exact value of the rebound velocity $V_0$ is not known, but is likely a fraction of the crater impact velocity $V_c$.
To investigate the effect of the rebound velocity, it was varied from $\unit [0.1]{m/s}$ to $\unit [1.0]{m/s}$.

The motion of each particle is governed by gravity, buoyancy, and viscous friction

\begin{equation}
\label{eg:Motion}
m \frac {d \vec{v}} {dt} = m\vec{g}\, \frac {\rho - \rho _w}{\rho _w} - m \frac {\vec{v}}{\tau_\mu},
\end{equation}
where $m$, $\vec{v}$, and $\rho$ are the particle mass, velocity and density, and $\tau_\mu$ is the characteristic viscous time

\begin{equation}
\label{eg:ViscousTime}
\tau_\mu = \frac {\rho d^2} {18 \mu},
\end{equation}
where $d$ is the particle's diameter and $\mu$ is the water viscosity. For $\rho = \unit [1.2]{g/cm^3}$, $d = \unit [350]{\mu m}$, and $\mu = \unit [10^{-3}] {Pa\cdot s}$,
$\tau_\mu = \unit [8.2]{ms}$. The short viscous time means that, after an initial impulse, a particle comes to a standstill in just a few milliseconds.
An approximate calculation of the distance traveled by a particle with an initial velocity of $\unit [0.5]{m/s}$ is $l  = \unit [4]{mm}$, 
suggesting that traveling from the cavity edge at $r = \unit [8]{cm}$ to the center requires at least 20 jolts 
or even more if only a fraction of the velocity is directed horizontally and gravity-induced deviations from horizontal motion are considered.

\begin{figure}[htbp]
\begin{center}
\includegraphics{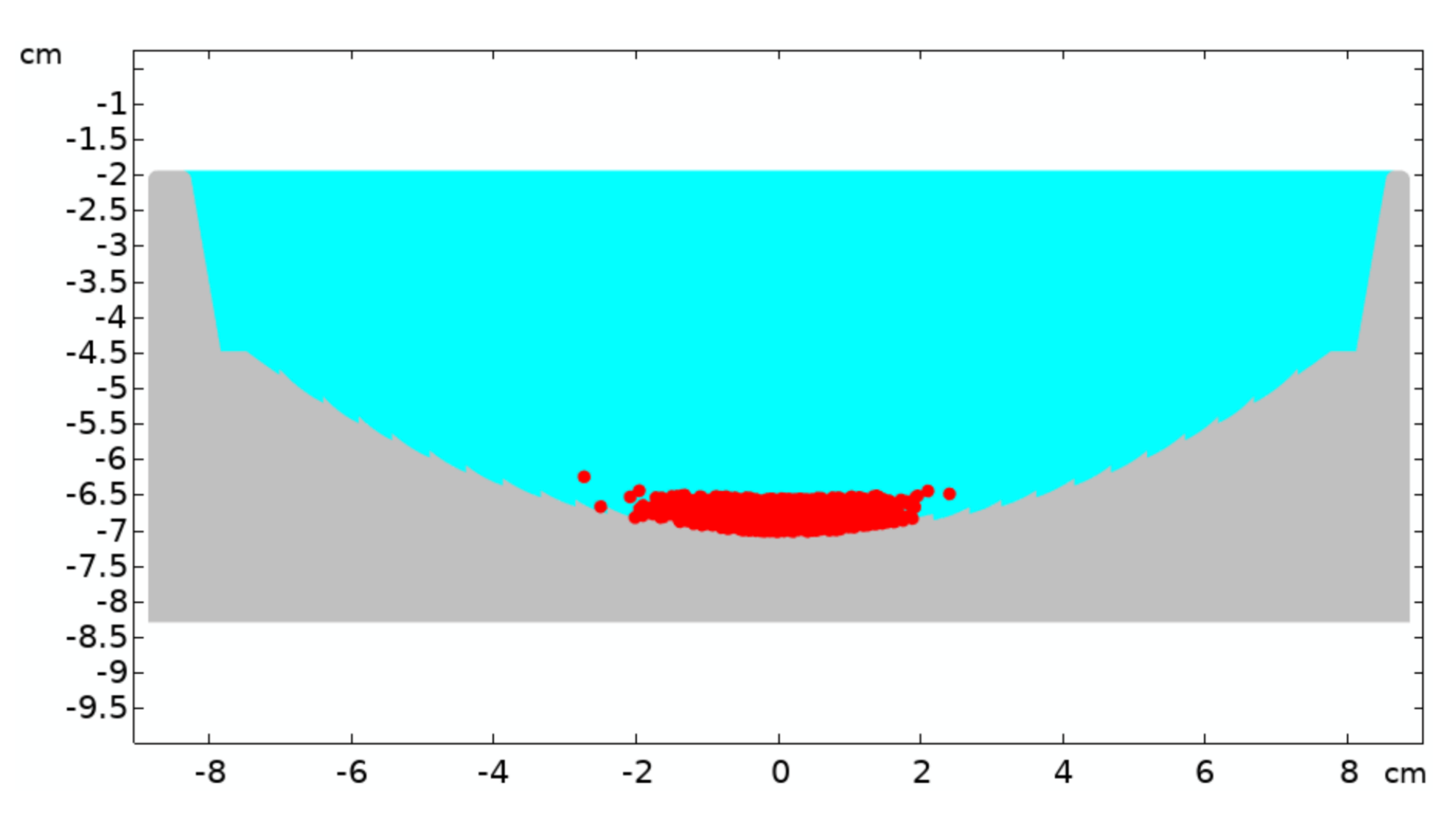} 
\caption{Particle distribution over the bottom of the cavity after 60 impacts at a rebound velocity of $\unit [0.5] {m/s}$.}
\label{fig:Simulations_T60}
\end{center}
\end{figure}

\begin{figure}[htbp]
\begin{center}
\includegraphics{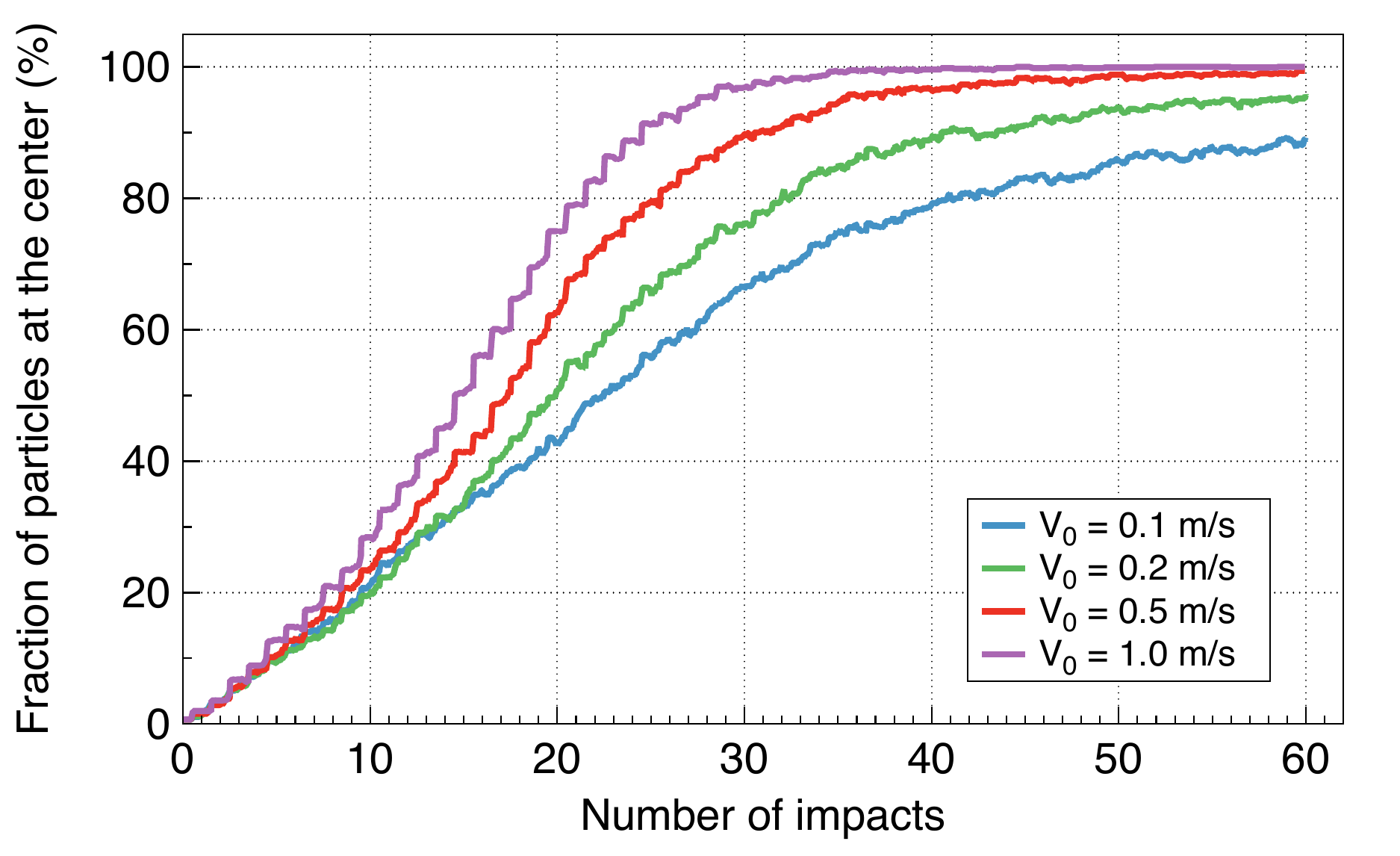} 
\caption{Evolution of particle concentration at the center at various rebound velocities $V_0$.
Shown is the time dependence of the fraction of the particle at the central area of the cavity $ r < \unit [2]{cm}$.
The small wiggles correspond to brief movements toward the center following each impact.}
\label{fig:ParticlesCenter}
\end{center}
\end{figure}

Figure~\ref{fig:Simulations_T60} shows the particle distribution over the bottom of the cavity after consecutive 60 impacts at a rebound velocity of $\unit [0.5] {m/s}$.
A clear concentration enhancement is visible in the central area of the cavity.
To explore the effect of the rebound velocity, it was varied from  $V_0 = \unit [0.1] {m/s }$ to $\unit [1.0] {m/s }$. 
The result shown in Fig.~\ref{fig:ParticlesCenter} confirms the quantitative estimates for the number of impacts needed.
Indeed, according to the simulations, collecting 70\% of all particles at a rebound velocity of $V_0 = \unit [0.5] {m/s}$ takes about 20 impacts, 
while reaching 90\% of all the particles requires approximately 30 impacts, which aligns well with the estimates.
Overall, the numerical results are in good agreement with the experimental data.

~

\section{Liposome protection against UV destruction in crater geometry}
\label{sec:UV_Crater}

\begin{figure}[htbp]
\begin{center}
\includegraphics{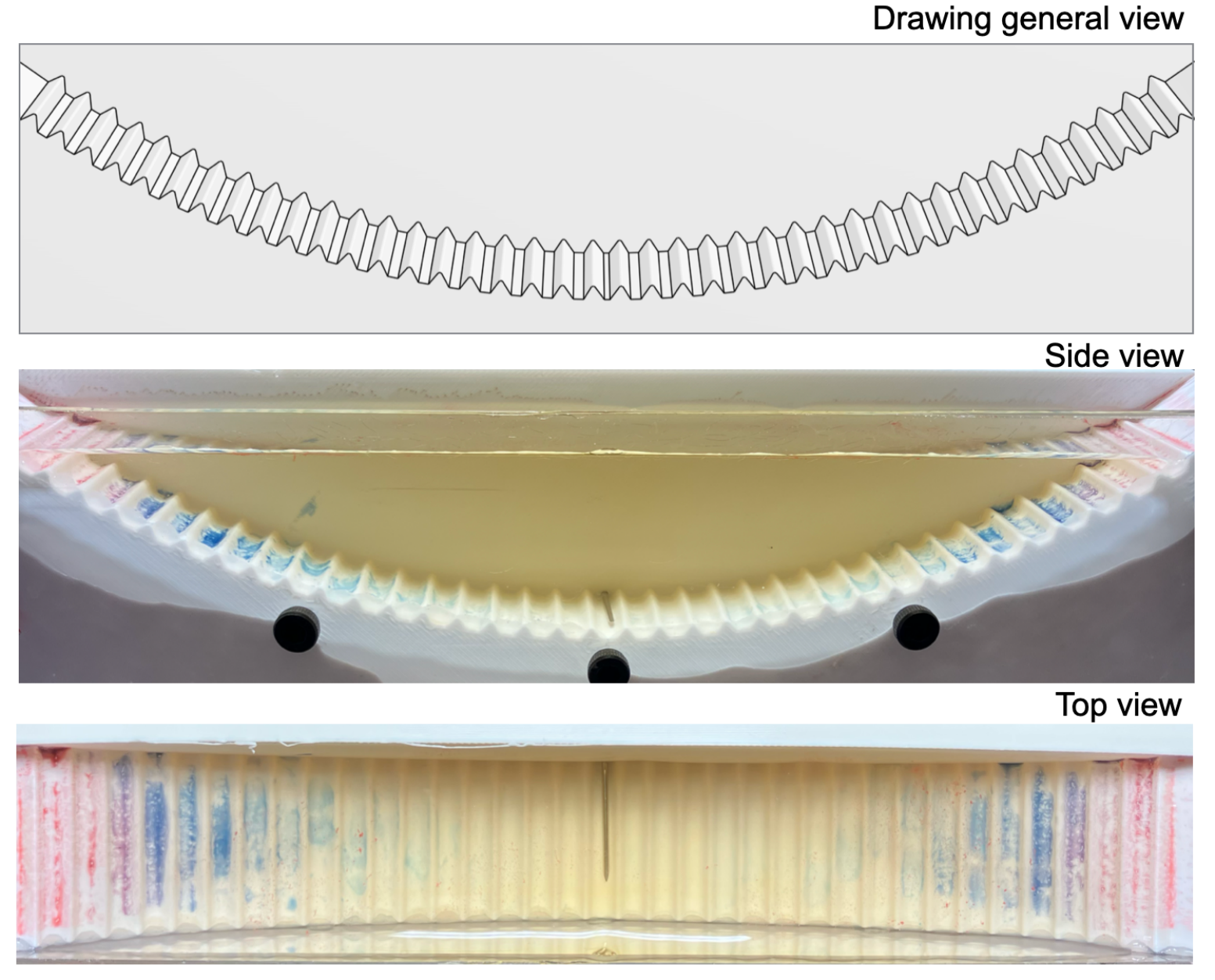} 
\caption{Illustration of liposome protection from UV. The top panel shows a drawing of the crater.
The side- and top-view photographs show the intensity distribution of the blue color after UV exposure.
When subjected to UV radiation, the liposomes go through a color change from the original white to blue, with the color intensity
correlated with the level of physical destruction of the liposomes.
Furthermore, the liposomes change to red when exposed to high heat.
The liposomes located at the edges and not covered by the solution turn to blue and then to red within the first few seconds of UV exposure,
likely due to the combined effects of UV and heat emitted from the lamp.
Liposomes covered by the solution demonstrate a gradual change of the blue color from deep-blue for those at the edge
to less intense and almost transparent, deeper toward the center.
}
\label{fig:Crater_UV}
\end{center}
\end{figure}

While the mechanism by which ferric salts protect liposomes from UV damage has been explained and quantified in our earlier publications~\cite{Subbotin2023AstrobiologyLipidWorld,Subbotin2023AstrobiologyUVProtection,Turner2025FrontiersUVLiposomes}, 
this section focuses on visually demonstrating this effect within the crater geometry.

We used large polymerizable UV-sensitive multilamellar liposomes with an average diameter  of $\unit [1.64]{\mu m}$
composed of Dibehenoyl-sn-glycero-3-phosphocholine (DBPC) and a 10,12-Pentacosadiynoic acid (PCDA) with a DBPC: PCDA molar ratio of 80:20, 
at a total lipid concentration of 7 mM (Encapsula NanoSciences, Brentwood, TN, USA). 
Liposomes were assembled in a sucrose solution, allowing them to achieve a specific density of $\unit [1.014] {g/cm^3}$.

The crater has a geometry very similar to that used in Section~\ref{sec:Setup}, except being fabricated in a planar geometry and with a transparent front wall.
The steps on the crater's floor are $\unit [2.5]{mm}$ high and spaced $\unit [5]{mm}$ apart.
Ten microliters of undiluted liposomes were placed on each as a strip. 
Then the crater was very slowly filled to a depth of $\unit [30]{mm}$ with a solution of iron trichloride at a concentration of $\unit [0.2]{g/L}$
and a specific density of $\unit [1.001]{g/cm^3}$ via a stainless steel tubing inserted in the lowest point.
The infusion speed was $\unit [1] {mL/min}$, and the total volume of solution was $\unit [94] {mL}$. 
The outermost steps were intentionally left uncovered.

A low-pressure UV bulb (Spectrolite, $\unit [15] {W}$, arc length $\unit [35] {cm}$) was positioned $\unit [30] {mm}$ above the crater. 
When subjected to UV radiation, the liposomes go through a color change from the original white to blue~\cite{Ann2009} with the color intensity
correlated with the level of physical destruction of liposomes, as was tested and confirmed by the liposome manufacturer.
Furthermore, the liposomes change to red when exposed to high heat.
The experiment shows that liposomes located at the edges and not covered by the solution turn to blue and then to red within the first few seconds of UV exposure,
likely due to the combined effects of UV and heat emitted from the lamp.
Liposomes covered by the solution demonstrate a gradual change of the blue color from deep-blue for those at the edge
to less intense and almost transparent, deeper toward the center.
Even after more than two hours of UV exposure, the liposomes at the lowest point of the crater's floor stayed intact.

\vspace {5mm}

\section{Summary and conclusions }
\label{sec:Summary} 

Various in-vitro studies have demonstrated that when liposomes come into physical contact, they tend to fuse, grow, exchange contents, and divide.
We hypothesize that gravity-driven and earthquake-assisted movement of liposomes within meteoric craters on Early Earth could lead to their accumulation at the crater floor. 
The paper presents both experimental and numerical investigations that demonstrate this phenomenon.

The ``crater'' for this experiment has a spherical cap shape with a sphere radius of $\unit [200]{mm}$ and a cap radius and height of $\unit [80]{mm}$ and $\unit [35]{mm}$, respectively,
and printed using water-tight 3D printing technology.
The base of the crater features ridges spaced every $\unit [5]{mm}$, each measuring $\unit [2.5]{mm}$ in height.
The crater is filled with water into which fluorescent red plastic ``liposomes'', with a diameter of $d = \unit [350]{\mu m}$ and density of $\rho = \unit [1.2]{g/cm^{3}}$ are submerged. 
Clearly, neither the size and shape of the ``crater" nor the size, material, and mass of the ``liposomes'' correspond to their real-life counterparts.
Rather, the purpose of this experiment is a demonstration of a general principle of liquid-submerged micro-particulate dynamics, driven
by gravitation and assisted by periodic disturbances.

Both experimental and numerical results show that applying a series of jolts with magnitudes similar to those of actual earthquakes causes the particle profile to gradually redistribute, eventually forming a density profile that peaks at the center.

\section*{Acknowledgments}
The authors thank Dr. Zahra Mirafzali (Encapsula NanoSciences, Brentwood, TN, USA) for technical support, for fulfilling our request to produce UV-sensitive liposomes, and for generously providing additional liposomes.


\bibliography{EarthquakesLiterature}

\end{document}